\documentclass{article}

\usepackage[english]{babel}

\usepackage[letterpaper,top=2cm,bottom=2cm,left=3cm,right=3cm,marginparwidth=1.75cm]{geometry}

\usepackage{amsmath}
\usepackage{graphicx}
\usepackage[colorlinks=true, allcolors=blue]{hyperref}

\usepackage{authblk} 

\title{\vspace{6em}Language learning shapes visual category-selectivity in deep neural networks}

\author[1, *]{Zitong Lu}
\author[2]{Yuxin Wang}

\affil[1]{MIT McGovern Institute for Brain Research}
\affil[2]{University of Cincinnati College of Medicine}

\affil[*]{corresponding author: zitonglu@mit.edu}

\date{} 

\begin{document}
\maketitle

\vspace*{2em}
\begin{abstract}
Category-selective regions in the human brain—such as the fusiform face area (FFA), extrastriate body area (EBA), parahippocampal place area (PPA), and visual word form area (VWFA)—support high-level visual recognition. Here, we investigate whether artificial neural networks (ANNs) exhibit analogous category-selective neurons and how these representations are shaped by language experience. Using an fMRI-inspired functional localizer approach, we identified face-, body-, place-, and word-selective neurons in deep networks presented with category images and scrambled controls. Both the purely visual ResNet and a linguistically supervised Lang-Learned ResNet contained category-selective neurons that increased in proportion across layers. However, compared to the vision-only model, the Lang-Learned ResNet showed a greater number but lower specificity of category-selective neurons, along with reduced spatial localization and attenuated activation strength—indicating a shift toward more distributed, semantically aligned coding. These effects were replicated in the large-scale vision-language model CLIP. Together, our findings reveal that language experience systematically reorganizes visual category representations in ANNs, providing a computational parallel to how linguistic context may shape categorical organization in the human brain.
\end{abstract}

\begin{quote}
\end{quote}

\newpage

\vspace*{1em}
\section{Introduction}

Category-selective regions in the human visual system, such as the Fusiform Face Area (FFA) ~\cite{Kanwisher1997, Kanwisher2006}, Extrastriate Body Area (EBA) ~\cite{Downing2001}, Parahippocampal Place Area (PPA) ~\cite{Epstein1998}, and Visual Word Form Area (VWFA) ~\cite{Dehaene2011, McCandliss2003}, play a crucial role in high-level visual processing. These areas exhibit hierarchical category selectivity, suggesting that categorical representations emerge progressively along the visual pathway ~\cite{Grill-Spector2014, Haxby2001}. In recent years, artificial neural networks (ANNs) have been widely used to model human visual processing ~\cite{Cadieu2014, Cichy2016, Kubilius2019, Yamins2014, Lu2023}. Studies have shown that ANNs can develop features resembling those in the biological visual system, such as low-level edge detectors and high-level semantic representations ~\cite{Yamins2016, Margalit2024, Lu2023d, Huang2021}. Importantly, ANNs not only approximate human visual behavior but also exhibit neural representations akin to those in the brain ~\cite{Cichy2016, Guclu2015, Kietzmann2019, Lu2023c, Lu2023e, Yamins2014, Yamins2016}. This has led to a “reverse engineering” approach, where neuroscientists analyze ANN representations to infer principles of human vision.

\vspace{1em}
Does ANNs naturally develop category-selective neurons and do their selectivity patterns align with those observed in the human brain? Recent work has examined category selectivity in ANNs from multiple perspectives on how such specialized responses emerge. Studies have reported face- and word-selective neurons in convolutional networks ~\cite{Baek2021, Xu2021a, Agrawal2024}, proving a foundation for exploring how high-level semantic properties may be encoded in these systems. Other studies analyzed specialization at the filter level for a small set of categories ~\cite{Dobs2022}, or introduced topographic constraints to explain structured selectivity ~\cite{Margalit2024, Blauch2022, Keller2021}. ~\cite{Prince2024} developed an analysis method for identifying category-selective neurons in self-supervised models. While these studies demonstrate that category selectivity can emerge across different models, they remain limited in scope, often focusing on few categories, lacking neuron-level rigorm, or avoiding comparisons across training objectives. 

\vspace{1em}
At the same time, while accumulating evidence suggests that language experience can modulate these visual representations (e.g., the emergence of the visual word form area), causal experiments on the role of language learning are not feasible in humans. This raises a critical question: does language learning influence the formation and distribution of category-selective neurons in visual systems? Does it increase their number or alter their selectivity patterns? Although language is known to play a role in concept learning ~\cite{Xu2002, Condry2008, Carruthers2002}, its impact on visual representations remains unclear.

\vspace{1em}
To address this question, we fine-tuned an ImageNet-pretrained ResNet to not only classify objects but also generate language embeddings, thereby forcing the model to integrate visual and linguistic information. Importantly, this “language-learned ResNet” was trained on the same ImageNet dataset, controlling for dataset differences when compared with the purely visual model. We then adopted a "functional localizer" approach, commonly used in fMRI research to identify category-selective regions in the brain ~\cite{Abassi2024, Li2024, Poldrack2007,Saxe2006,Kanwisher1999,Duncan2009}, presenting images from different categories (faces, bodies, places, words, scrambled images) and identifying category-selective neurons using statistical criteria. Our results show that category-selective neurons emerge across multiple layers of ANNs, increasing in proportion at deeper levels. Moreover, Lang-learned ResNet exhibits a higher proportion of category-selective neurons than the baseline ResNet, but they are less selective and more uniformly distributed. These findings suggest that language learning increases the number of category-selective neurons while reducing their selectivity strength, shedding light on how multimodal learning reshapes visual representations.

\vspace{1em}
The key contributions of our current study are as follows: (1) We retrained a highly controlled language-learned model that allows fair comparison with a purely visual model to explore how language learning would shape internal representations. (2) We systematically investigated category-selective neurons in both purely visual and controlled language-learned ANNs across hierarchical layers. (3) We reveal the influence of language on category selectivity in neural representations and offer new insights into ANNs as models of human vision and multimodal learning.

\section{Results}

\begin{figure}[htbp!]
\centering
\includegraphics[width=1\textwidth]{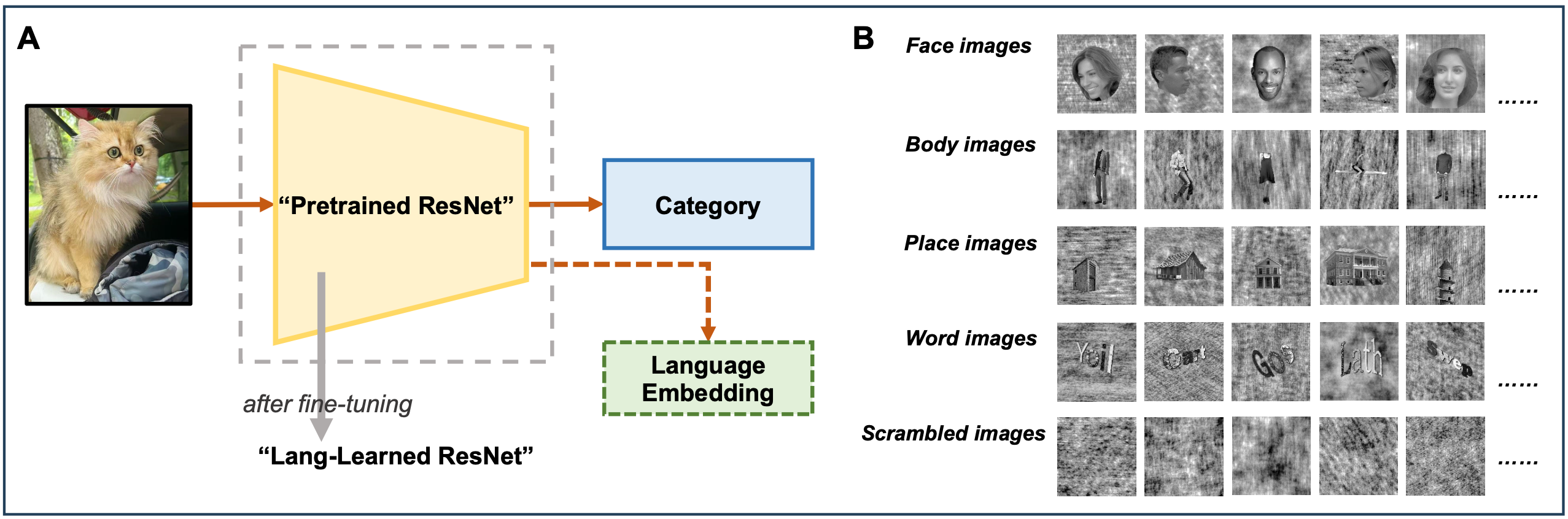}
\caption{\label{Figure1}Lang-Learned ResNet and stimuli in our study. (A) We fine-tuned pretrained ResNet by adding the language embedding generation task to the original model architecture on ImageNet to obtain Lang-Leanred ResNet. (B) Image stimuli used in the functional localizer experiment. Images were divided into five categories: face, body, place, word, and scrambled images.}
\end{figure}

To examine how language learning influences category-selective representations in ANNs, we trained a Lang-Learned ResNet - a ResNet-50 model fine-tuned with an additional language embedding generation objective (Figure \ref{Figure1}A; see details in Materials and Methods section) - alongside a purely visual ResNet-50 as a structural control. Mirroring the logic of functional localizer studies in human neuroimgaing, we presented each model with images from five categories (faces, bodies, places, words, and scrambled controls; Figure \ref{Figure1}B) to identify neurons selectively responsive to specific regions. The selectivity of these neurons was then tested using held-out image sets, ensuring non-circular validation (see details in Materials and Methods section). This ANN-based localizer framework allows us to directly quantify how category selectivity emerges across layers and to evaluate how additional language supervision reshapes these categorical representations. By adopting a paradigm conceptually aligned with fMRI and intracranial studies, this analysis bridges representational principles between artificial and biological vision systems, offering insights into how language experience can reorganize visual feature tuning.

\subsection{Language learning attenuates the activation strength of category-selective neurons}

\begin{figure}[t!]
\centering
\includegraphics[width=1\textwidth]{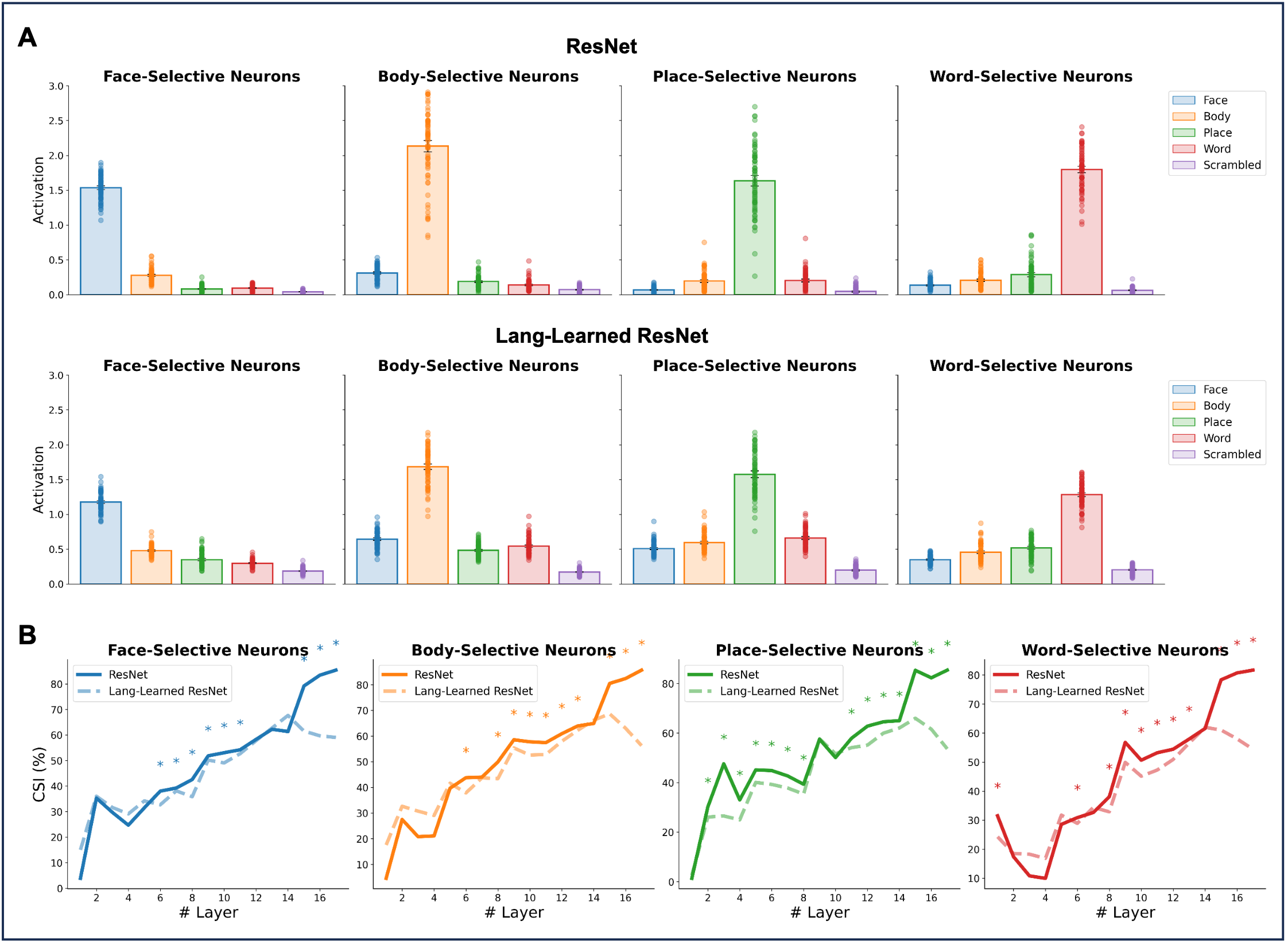}
\caption{\label{Figure2}Category-selective activations in ResNet and Lang-Learned ResNet. (A) Response profiles of category-selective neurons from "layer4.2.relu". The top raw shows results for ResNet, and the bottom row for Lang-Learned ResNet. Each individual dot corresponds to a image. (B) Category selectivity index (CSI) across layers. From left to right, each panel shows the proportion of neurons selective for faces (blue), bodies (orange), places (green), and words (red) at each layer of ResNet (solid lines) and Lang-Learned ResNet (dashed lines). Asterisk indicates CSI of ResNet is significantly greater than CSI of Lang-Learned ResNet, $p<.05$.}
\end{figure}

We first examined how language learning influences the activation of category-selective neurons. To begin, we focused on a single representative layer within the ResNet-50 architecture - "layer4.2.relu" - to analyze activation patterns at a late stage of visual processing. Statistical analyses across all neurons in this layer revealed that both models exhibited category-selective neurons for faces, bodies, places, and words. Figure \ref{Figure2}A illustrates the activation strengths of these category-selective neurons across different image categories.

\vspace{1em}
To extend this analysis beyond a single layer, we systematically quantified activation selectivity across all layers using the category selectivity index (CSI) (Figure \ref{Figure2}B). Across the network hierarchy, ResNet consistently exhibited higher CSI values than Lang-Learned ResNet, particularly in later layers that are more involved in higher-level visual representations. This pattern indicates that language learning leads to an attenuation in the activation strength of category-selective neurons, suggesting that the incorporation of linguistic supervision may compress or regularize visual feature tuning, especially in higher-level layers.

\vspace{1em}

\subsection{Language learning increases the number of category-selective neurons}

\begin{figure}[htbp!]
\centering
\includegraphics[width=1\textwidth]{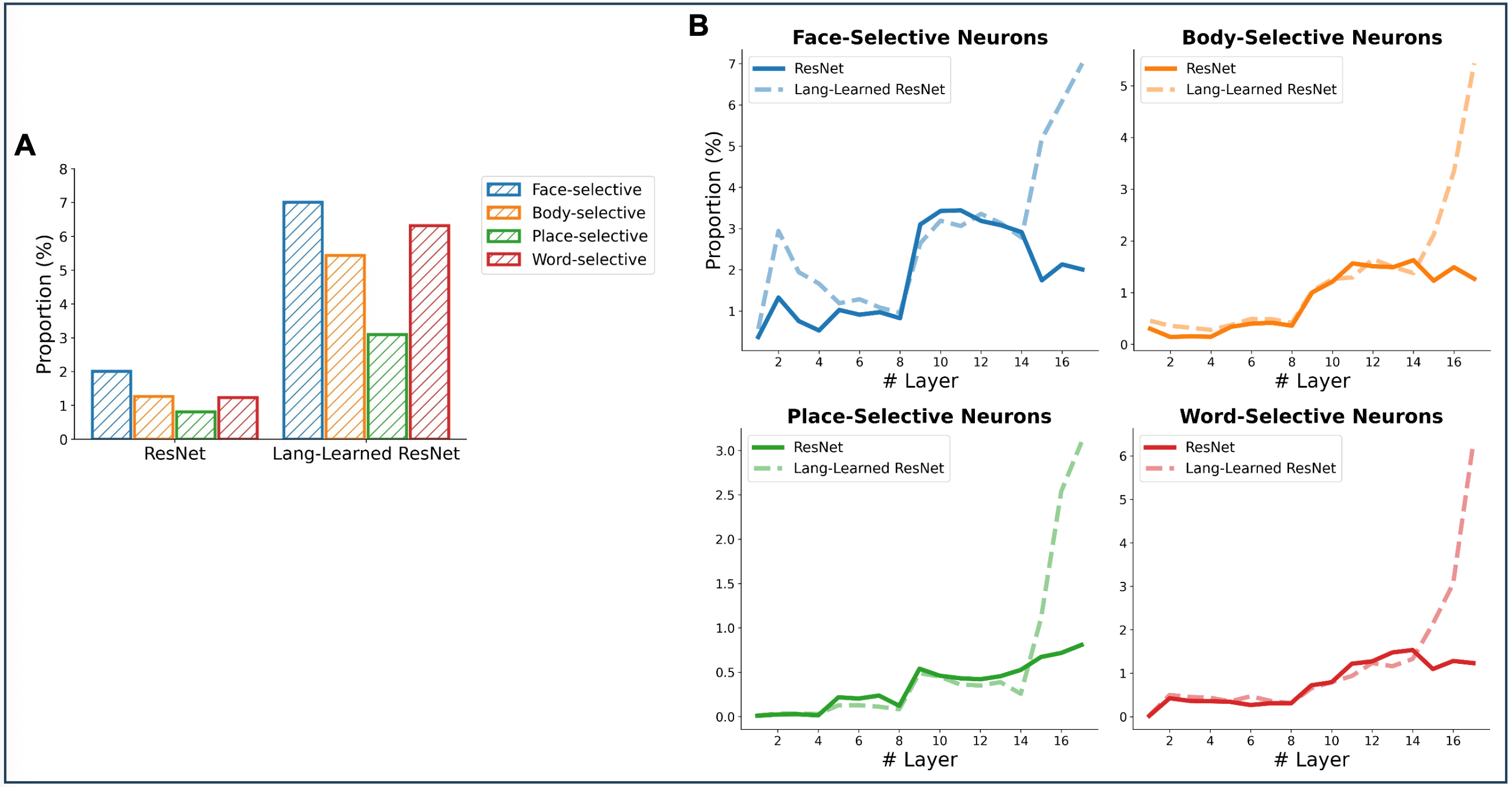}
\caption{\label{Figure3}Proportion of Category-selective neurons in ResNet and Lang-Learned ResNet. (A) Proportion of category-selective neurons in "layer4.2.relu". Neurons were classified as category-selective if they exhibited significantly higher activation for one category compared to all others. (B) Proportion of category-selective neurons across layers.}
\end{figure}

Beyond activation strength, we next examined whether language learning alters the number of category-selective neurons. In "layer4.2.relu" of ResNet, we observed the highest proportion of face-selective neurons (2.01$\%$), followed by body-selective neurons (1.27$\%$), word-selective neurons (1.23$\%$), and place-selective neurons (0.80$\%$). In contrast, the corresponding layer in Lang-Learned ResNet showed substantially higher proportions of category-selective neurons across all categories — with 7.01$\%$ for faces, 5.34$\%$ for bodies, 3.09$\%$ for places, and 6.32$\%$ for words (Figure \ref{Figure3}A).

\vspace{1em}
To extend this analysis across the network hierarchy, we quantified  the layer-wise progression of category-selective neuron proportions (Figure \ref{Figure3}B). In the purely visual ResNet, the proportion  of category-selective neurons initially increased with depth and then declined in later layers. In contrast, Lang-Learned ResNet displayed a monotonically increasing trend, maintaining a higher proportion of category-selective neurons in later layers than ResNet.

\vspace{1em}
Overall, while the purely visual ResNet contained fewer category-selective neurons, those neurons exhibited stronger and more specific selectivity. Lang-Learned ResNet, by contrast, recruited a larger population of neurons showing category preference but with broader, less selective tuning. This dissociation - a greater number of category-selective neurons but lower CSI values - suggests that language learning broadens categorical tuning and distributes category information across a wider neural population, potentially reflecting a more generalized representational organization induced by linguistic supervision.

\vspace{1em}

\subsection{Language learning reduces layer-wise feature-map variation among category-selective neurons}

To further investigate how category-selective neurons are distributed across the spatial layout of convolutional feature maps, we examined the spatial organization of these neurons and compared their distributional patterns between models. Specifically, we asked whether category-selective neurons are spatially localized or more evenly distributed across the feature maps.

\begin{figure}[t!]
\centering
\includegraphics[width=1\textwidth]{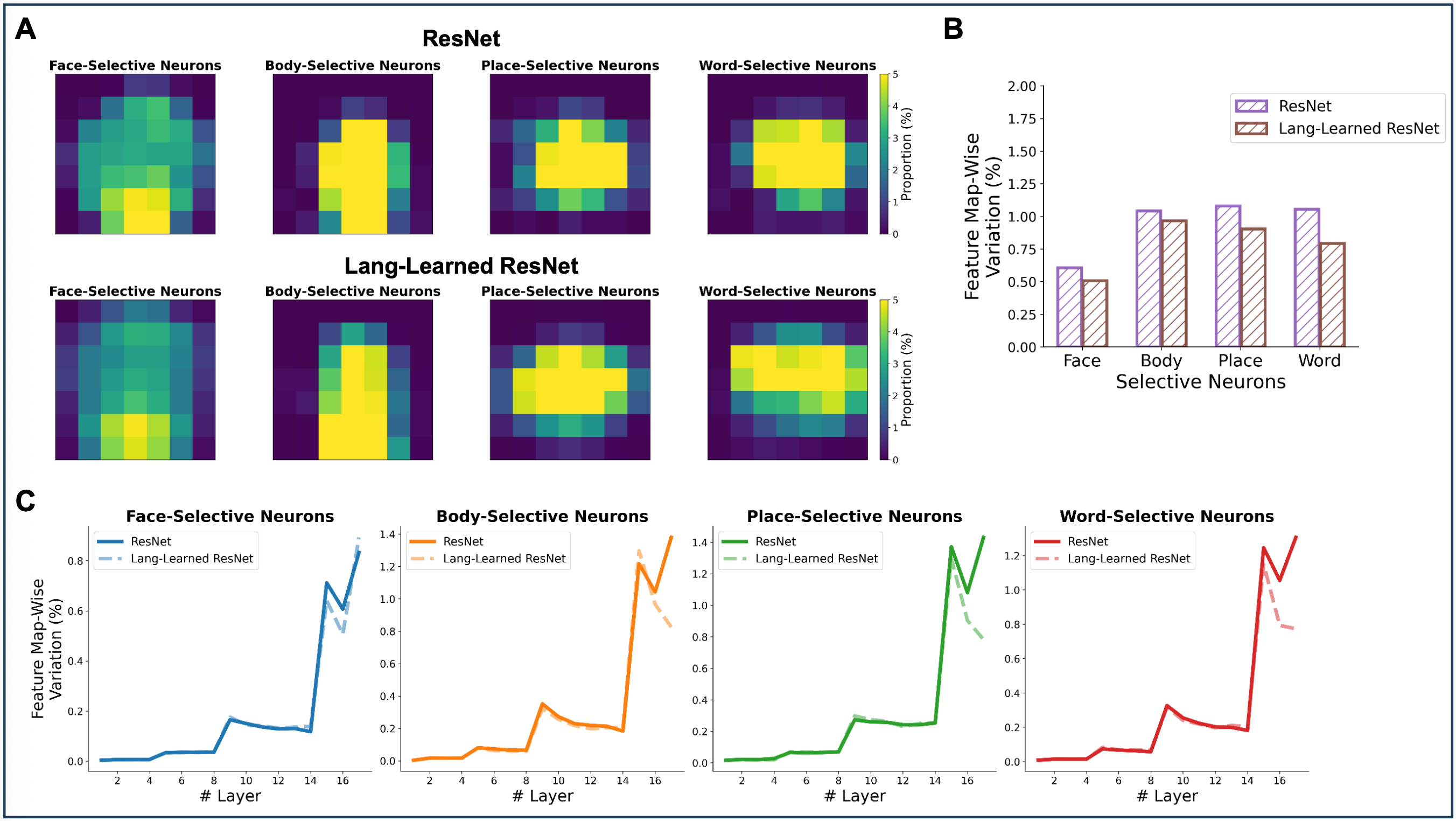}
\caption{\label{Figure4}Spatial distribution of category-selective neurons in ResNet and Lang-Learned ResNet. (A) Feature map-wise distribution of category-selective neurons. Each heatmap represents the spatial distribution of face-, body-, place-, and word-selective neurons across the feature map in "layer4.2.relu" for ResNet (top raw) and Lang-Learned ResNet (bottom row). The color scale indicates the proportion of neurons at each spatial location in the feature map. (B) Quantification of feature map-wise variation in category-selective neuron distribution. (C) Feature map-wise variation in category-selective neurons across layers.}
\end{figure}

\vspace{1em}
We first focused on the "layer4.2.relu" layer. All channels in this layer were aggregated to visualize the cumulative occurrence ratio of category-selective neurons at each spatial position in the feature map, enableing a direct comparison between ResNet and Lang-Learned ResNet. Figure \ref{Figure4}A shows the spatial proportion of different category-selective neurons (i.e., the number of selective neurons appearing at a given location divided by the total number of channels). Both models exhibited location-specific preferences for different category-selective neurons, rather than a uniform distribution across locations. To quantify spatial distribution patterns, we computed the variance of category-selective neuron proportions across spatial positions. Lower variance indicates a more even (less localized) distribution. As shown in Figure \ref{Figure4}B, ResNet exhibited substantially higher variance, indicating that its category-selective neurons were more spatially localized, whereas the Lang-learned ResNet showed a more uniform spatial distribution across the feature map.

\vspace{1em}

\vspace{1em}
Extending this analysis across all layers, we found that the reduced spatial variation in Lang-learned ResNet was predominantly observed in later layers, whereas early and mid-level layers showed comparable variation between models (Figure \ref{Figure4}C). Together, these results suggest that while the purely visual ResNet favors more localized feature selectivity, language learning promotes broader and more globally distributed representations, potentially facilitating a more holistic integration of category information in higher-level processing stages.

\subsection{CLIP recapitulates the language-learning effects observed in Lang-Learned ResNet}

Finally, to test whether the effects of language learning generalize to more commonly used multimodal models, we examined CLIP—a ResNet-50–based architecture trained with contrastive image–text supervision. Although our Lang-learned ResNet provides a more controlled comparison with the purely visual model, CLIP offers an opportunity to evaluate whether similar representational changes emerge under large-scale natural language supervision.

\begin{figure}[t!]
\centering
\includegraphics[width=1\textwidth]{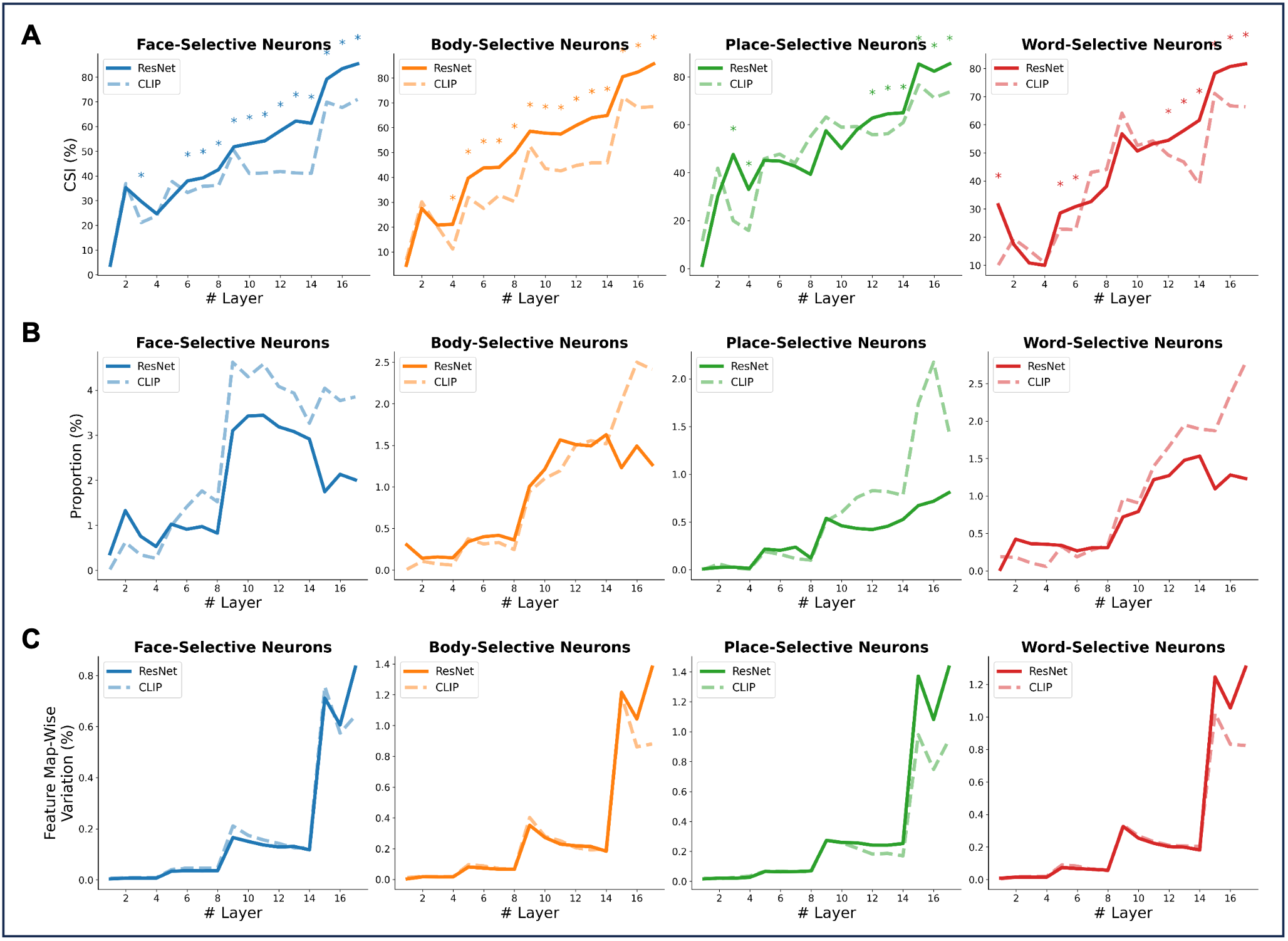}
\caption{\label{Figure5}Category-selectivity between ResNet and CLIP. (A) Category selectivity index (CSI) across layers. Asterisk indicates CSI of ResNet is significantly greater than CSI of CLIP, $p<.05$. (B) Proportion of category-selective neurons across layers. (C) Feature map-wise variation in category-selective neurons across layers. From left to right, each panel shows the proportion of neurons selective for faces (blue), bodies (orange), places (green), and words (red) at each layer of ResNet (solid lines) and Lang-Learned ResNet (dashed lines).}
\end{figure}

\vspace{1em}
As shown in Figure \ref{Figure5}, CLIP exhibited patterns that closely paralleled those of Lang-Learned ResNet. First, CLIP showed attenuated category-selective activations, with lower CSI values than the purely visual ResNet across higher layers (Figure \ref{Figure5}A). Second, CLIP displayed an increased proportion of category-selective neurons relative to ResNet (Figure \ref{Figure5}B), indicating broader recruitment of units responsive to categorical information. Third, CLIP demonstrated reduced layer-wise feature-map variation among category-selective neurons (Figure \ref{Figure5}C), suggesting a more spatially uniform and distributed organization.

\vspace{1em}
Together, these results reveal that the key effects of language learning—attenuated activation strength, expanded category-selective populations, and reduced spatial variation—are reproducible in CLIP. This convergence supports the view that linguistic supervision systematically reorganizes category representations in convolutional networks, promoting more distributed and semantically aligned visual encoding.

\vspace{1em}

\section{Discussion}

Through a systematic analysis of neuronal activity in ANNs, we demonstrate that these models contain category-selective neurons analogous to those found in the human brain, such as face-selective neurons (akin to FFA), body-selective neurons (akin to EBA), place-selective neurons (akin to PPA), and word-selective neurons (akin to VWFA). This finding indicates that current ANN models exhibit category-selective responses resembling those observed in biological visual systems. Our results align with prior studies reporting the emergence of category-selective neurons in convolutional networks studies \cite{Margalit2024, Dobs2022, Blauch2022, Prince2024, Keller2021}. Importantly, our approach mirrors human fMRI functional localizer studies by incorporating scrambled images as better control conditions to rule out low- and mid-level visual feature-driven selectivity. This design ensures that the observed category selectivity genuinely reflects semantic processing rather than basic visual properties. Notably, Notably, none of the examined models were explicitly trained to recognize these categories—ImageNet-trained ResNet, for example, was never exposed to face or word classification. Yet, category-selective neurons still emerged, particularly for words, demonstrating that such representations can arise spontaneously from large-scale visual learning, without direct category-level supervision.

\vspace{1em}
How does language learning influence category-selective neurons? Humans cannot be experimentally "deprived" of language to assess its influence on visual category selectivity. To approximate this contrast computationally, we designed the Lang-Learned ResNet as a linguistically experienced model, while the purely visual ResNet serves as a language-deprived control. Both models share identical architectures and are trained on the same ImageNet images, but the Lang-Learned ResNet includes an additional language embedding generation objective that introduces multimodal supervision analogous to language experience. This design allows us to isolate the effect of linguistic learning on category-selective representations within a controlled framework.

\vspace{1em}
Our findings show that the Lang-Learned ResNet exhibits a greater number of category-selective neurons but lower category selectivity index (CSI) compared to the vision-only ResNet. This pattern indicates that neurons in the Lang-Learned ResNet are more broadly tuned—responding to multiple semantically related categories—whereas neurons in ResNet exhibit sharper, more exclusive selectivity. We interpret this as evidence that language experience broadens categorical representations while reducing single-unit specificity, consistent with the idea that linguistic context enhances semantic associations between categories.

\vspace{1em}
Further, the spatial distribution of these neurons differed between models. In ResNet, category-selective neurons were spatially localized within specific feature-map regions, indicating reliance on localized visual features for discrimination. In contrast, Lang-Learned ResNet showed a more even spatial distribution and lower layer-wise variance, implying that language supervision promotes globally distributed representations. This broader, more integrated encoding may facilitate holistic category understanding and improved generalization — consistent with language’s role in linking semantically related visual concepts.

\vspace{1em}
To examine whether these language-learning effects extend to a widely adopted multimodal framework, we tested CLIP, which shares a ResNet-50 backbone but is trained via large-scale contrastive image–text alignment. Strikingly, CLIP exhibited the same pattern as Lang-Learned ResNet: attenuated category-selective activation strength, increased proportions of category-selective neurons, and reduced spatial variation across layers. These convergent results suggest that language supervision—whether introduced through explicit embedding regression or large-scale contrastive learning—consistently reorganizes category representations. The replication of these patterns in CLIP reinforces our interpretation that the Lang-Learned ResNet serves as a human-relevant model of the linguistically experienced brain, whereas the vision-only ResNet represents a visual system deprived of linguistic grounding.

\vspace{1em}
While earlier studies have identified category-selective neurons in CNNs \cite{Agrawal2024, Baek2021, Xu2021a, Margalit2024, Dobs2022, Blauch2022, Prince2024, Keller2021}, our study represents, to our knowledge, the first systematic investigation comparing neuron-level category selectivity for face, body, place, and word representations between structurally matched vision-only and vision+language models. Beyond confirming the existence of category-selective neurons, we reveal how language supervision modulates their distribution, selectivity strength (CSI), and spatial organization. Our study suggest novel insights into: (1) The hierarchical emergence of category-selective neurons in ANNs, mirroring the increasing category selectivity observed in biological vision. (2) The role of language learning in expanding category-selective neurons while reducing their specificity, suggesting that language enhances category relationships. (3) The greater spatial uniformity and cross-layer stability of category-selective neurons in Lang-Learned ResNet compared to ResNet, highlighting the effect of language in promoting global category representations. These findings contribute to a deeper understanding of ANNs as models of human vision and provide new insights into how multimodal learning reshapes visual representations. 

\vspace{1em}
Beyond quantifying selectivity differences, our findings have broader functional and theoretical implications. The vision-only ResNet favors sparse, localized, and highly specific feature detectors, optimizing fine-grained visual discrimination but limiting semantic generalization. In contrast, the Lang-Learned ResNet and CLIP show a shift toward more distributed, population-level representations that integrate visual and semantic regularities. This transformation—from local feature-based to global semantic-based organization—may confer computational advantages for multimodal reasoning, robustness, and abstraction. Biologically, this duality mirrors the human visual system’s representational hierarchy: specialized cortical regions (e.g., FFA, PPA, VWFA) coexist with distributed semantic representations that integrate visual knowledge with linguistic and conceptual context. Language experience thus appears to “semanticize” visual representations, aligning ANN transformations with principles of human visual cognition.

\vspace{1em}
Several open questions remain: First, our study focuses on trained models, rather than examining category-selective neurons in randomly initialized networks. Some studies have suggested using untrained networks as an analogy for the “infant brain,” \cite{Baek2021, Kim2021, Zhou2022} but we argue that this analogy is flawed since infant brains already possess structured neural connections at birth. Future work should investigate how category selectivity emerges dynamically over learning stages, paralleling developmental visual learning in humans. Second, while our study reveals category-selective neurons in ANNs, it does not directly compare them to category-selective regions in the human brain (e.g., FFA, PPA, VWFA). Future research could leverage fMRI or ECoG data to compare category-selective activations in ANN neurons and human cortical regions, further validating the biological plausibility of ANN representations. Third, our study characterizes the distribution of category-selective neurons but does not directly test their computational importance. causal analyses—such as ablating category-selective neurons—could determine their functional contributions to recognition and generalization. Fourth, the ANNs in our study were trained on general object classification tasks (ImageNet for ResNet and contrastive learning for CLIP). extending beyond classic object categories, future work should explore how language modulates other conceptual dimensions—such as real-world size, animacy, food-relatedness, and material properties \cite{Bao2020, Coggan2023, Huang2022, Jain2023, Khaligh-Razavi2018, Khosla2022, Konkle2012a, Konkle2013, Lu2023d} — to test whether language-induced semanticization generalizes to higher-order visual concepts. Recent studies suggest that models optimized using neural data exhibit stronger food-related representations than those trained solely on images \cite{Lu2024a, Lu2024b}. Future work should explore how these categorical features are encoded differently in ANNs and the human brain, and how language learning influences their representation.

\vspace{1em}
In summary, by integrating fMRI-inspired functional localization with neuron-level analysis, we show that category-selective neurons naturally emerge in ANNs and that language experience systematically reshapes their organization. Language learning increases the prevalence of category-selective neurons but reduces their single-unit specificity and spatial localization, producing more globally distributed, semantically organized representations. The replication of these effects in CLIP further demonstrates that language supervision provides a consistent and biologically meaningful organizing principle for visual representations. Together, these findings suggest that multimodal learning transforms vision from a feature-based to a semantic-based system—offering a computational bridge between artificial and biological vision.

\vspace{2em}
\section{Materials and Methods}

In this section, we describe the selection of models, including the training steps of Lang-learned ResNet and the rationale behind it, the stimuli used for our functional localizer experiments in ANNs, the method for identifying category selective neurons, and the metrics used to quantify selectivity. Additional methodological details are provided in the corresponding result section.

\subsection{Model selection}

To investigate category-selective neurons in deep neural networks, we compared a purely visual model (ResNet-50) with a fine-tuning language-learned ResNet-50. Additionally, we also analyized ResNet-50-based CLIP as a second comparison. We chose these ResNet-structured models for the following reasons: (1) CNN-based architectures better align with human vision: Prior research suggests that convolutional neural networks (CNNs) are more biologically plausible than vision transformers (ViTs). CNNs exhibit a hierarchical organization similar to the human visual cortex, have higher neural and behavioral similarity to humans \cite{Geirhos2021}, and demonstrate human-like adversarial vulnerability \cite{Bhojanapalli2021}. Given these advantages, we chose ResNet over ViTs. CNNs inherently incorporate spatial and hierarchical inductive biases—such as localized receptive fields, weight sharing, and spatial pooling—that closely mirror the organization of the human visual cortex. These properties enable CNNs to effectively capture low-level visual features (e.g., edges and textures) and to progressively build more complex representations, much like the hierarchical processing observed in early and intermediate visual areas. In contrast, while transformer models have recently shown promise in modeling high-level semantic information and behavioral responses, their lack of built-in spatial constraints and inductive biases makes them less naturally aligned with the retinotopic and localized processing characteristic of human vision. Thus, despite emerging evidence that transformers can also approximate aspects of neural data, our focus on CNN-based models is motivated by their robust performance in modeling detailed, fine-grained visual processing—a hallmark of biological vision systems. (2) Structural control between models: To compare purely visual models with vision-language models, we controlled for architecture by retrained our Lang-learned ResNet and selecting ResNet-50-based CLIP, which maintains the same ResNet-50 backbone. This ensures that any observed differences between models are primarily due to language supervision rather than architectural discrepancies.

\vspace{1em}
\textbf{ResNet}: We used pretrained ResNet-50 on ImageNet \cite{Deng2009}. The model is trained to classify images into 1,000 object categories using a cross-entropy loss function. Training was performed with stochastic gradient descent (SGD) with momentum, incorporating extensive data augmentation techniques such as random cropping, horizontal flipping, and normalization. These training practices ensure that the model learns robust, invariant visual features that are representative of a purely vision-based approach.

\vspace{1em}
\textbf{Lang-learned ResNet}: To construct a language-learned version of ResNet, we fine-tuned a Lang-Learned ResNet as a strictly controlled comparison model (Figure \ref{Figure1}A). The training procedure was as follows: based on a pretrained ResNet-50, we added an additional language embedding generation task. Specifically, the model's final layer was required not only to perform image category classification but also to generate a corresponding language embedding for each image input. For supervision, we first used BLIP-2 to generate a caption for every ImageNet image, and then extracted a 768-dimensional language embedding for each caption using MPNet-v2. During training, the model jointly optimized two loss terms:

\begin{equation}
    \mathcal{L} = \mathcal{L}^{Classification} + \beta \cdot \mathcal{L}^{LangEmbedding}
\end{equation}

where $\mathcal{L}^{Classification}$ is computed using cross-entropy loss, and $\mathcal{L}^{LangEmbedding}$ is computed using mean-squared error (MSE) between the predicted and target language embeddings. The weighting coefficient $\beta$ was fixed at 100 to ensure comparable magnitudes between the two loss components. The model was optimized with the Adam optimizer (learning rate = .00001) until convergence, defined as no further decrease in the overall loss. Lang-Learned ResNet shows a similar higher classification performance on ImageNet (Top-1: 76.1245\%, Top-5: 92.9616\%) to purely visual ResNet (Top-1: 76.1328\%, Top-5: 92.8679\%).

\vspace{1em}
Therefore, both ResNet and Lang-Learned ResNet share identical architectures and were trained solely on ImageNet. The only difference is that Lang-Learned ResNet was additionally trained with a language embedding generation objective, allowing subsequent experiments to isolate and quantify the specific influence of language learning on network representations.

\vspace{1em}
\textbf{CLIP}: We also conducted our experiments on pretrained ResNet-50-based CLIP from OpenAI, which has the same convolutional backbone but is trained using a contrastive learning objective with paired image-text data, as described in \cite{Radford2021}. CLIP’s training involves pairing images with corresponding text descriptions to learn a joint multimodal representation space. The contrastive loss is designed to align the representations of matching image-text pairs while pushing apart non-corresponding pairs. This learning paradigm benefits from the rich semantic information provided by textual annotations, which augments visual feature learning. Similar to the ResNet-50 model, CLIP’s training incorporates data augmentation and normalization techniques, ensuring robust feature extraction.

\vspace{1em}
Additionally, we focused our analysis on all layers following ReLU activations in ResNet-50, totaling 17 layers ("relu", "layer1.0.relu", "layer1.1.relu", "layer1.2.relu", "layer2.0.relu", "layer2.1.relu", "layer2.2.relu", "layer2.3.relu", "layer3.0.relu", "layer3.1.relu", "layer3.2.relu", "layer3.3.relu", "layer3.4.relu", "layer3.5.relu", "layer4.0.relu", "layer4.1.relu", "layer4.2.relu", and "avgpool"). This choice was motivated by two key considerations: First, ReLU introduces sparsity by zeroing out negative activations, making feature representations more biologically plausible. Second, nonlinear feature transformations primarily occur after ReLU, meaning category selectivity is more meaningfully assessed at these points.

\subsection{Stimuli}

We selected 144 images each from four categories: face, body, place, and word, sourced from the fLoc functional localizer package \cite{Stigliani2015} (Figure \ref{Figure1}B). However, when investigating category-selective neurons, using only these four categories could raise concerns regarding whether the selectivity is truly driven by high-level semantic information or merely by low-level visual features such as edges, textures, and spatial frequency.

\vspace{1em}
To control for low-level feature-driven selectivity, we also included 144 scrambled images as control conditions (Figure \ref{Figure1}B). This manipulation ensures that any observed category selectivity is more likely to reflect high-level semantic processing rather than simple visual properties. Our stimulus selection and design follow rigorous protocols from fMRI studies, ensuring that our analysis of category-selective neurons in ANNs aligns with established methodologies used in human neuroscience research.

\vspace{1em}
Importantly, to further avoid circular validation, we randomly split the 144 images in each category into two equal sets of 72. The first set was used to conduct the "functional localizer" to identify category-selective neurons, while the second set was used to test the category selectivity of the identified neurons and for subsequent analyses.

\subsection{"Functional localizer" in ANNs}

Similar to functional localizer tasks in fMRI, we presented the 360 images (72 images per category) to each ANN model, recording the activation of every neuron in response to different images from different categories. A neuron was identified as category-selective if its activation was significantly stronger for one category compared to all others. Specifically, for each neuron, we conducted independent t-tests comparing activations between images from one category and each of the other categories. For instance, to be classified as a face-selective neuron, a neuron needed to show: Face$>$Body and Face$>$Scene and Face$>$Word and Face$>$Scrambled Scene and Face$>$Scrambled Word, with p$<$0.05 for each comparison. This allowed us to quantify the proportion of category-selective neurons at each layer and measure the activation strength of these neurons across different visual stimuli.

\vspace{1em}
Then we confirmed the category-selectivity of these neurons on the held-out 360 images (72 images per category) on each ANN model. Also, the subsequent analyses were conducted using these held-out 360 images rather than the 360 images used for the localizer.

\subsection{Category selectivity index (CSI)}

Since activation magnitudes vary across layers, we computed a Category Selectivity Index (CSI) to quantify category selectivity in a layer-independent manner:
\begin{equation}
CSI = \frac{R_{\text{preferred}} - R_{\text{non-preferred}}}{R_{\text{preferred}} + R_{\text{non-preferred}}}\times100\%
\end{equation}
where $R_{\text{preferred}}$ is the mean activation for images of the neuron’s preferred category, and $R_{\text{non-preferred}}$ is the mean activation across all other categories. A higher CSI indicates stronger category selectivity. A lower CSI suggests weaker selectivity or broad tuning across categories. By comparing CSI across layers and models, we could assess how category selectivity evolves through hierarchical processing and how language supervision reshapes category representations.

\vspace{2em}
\bibliographystyle{unsrt}
\bibliography{sample}

\begin{thebibliography}{10}

\bibitem{Kanwisher1997}
Nancy Kanwisher, Josh McDermott, and Marvin~M. Chun.
\newblock {The Fusiform Face Area: A Module in Human Extrastriate Cortex Specialized for Face Perception}.
\newblock {\em Journal of Neuroscience}, 17(11):4302--4311, jun 1997.

\bibitem{Kanwisher2006}
Nancy Kanwisher and Galit Yovel.
\newblock {The fusiform face area: a cortical region specialized for the perception of faces}.
\newblock {\em Philosophical Transactions of the Royal Society B: Biological Sciences}, 361(1476):2109--2128, 2006.

\bibitem{Downing2001}
P.~E. Downing, Y.~Jiang, M.~Shuman, and N.~Kanwisher.
\newblock {A Cortical Area Selective for Visual Processing of the Human Body}.
\newblock {\em Science}, 293(5539):2470--2473, 2001.

\bibitem{Epstein1998}
Russell Epstein and Nancy Kanwisher.
\newblock {A cortical representation of the local visual environment}.
\newblock {\em Nature}, 392(6676):598--601, 1998.

\bibitem{Dehaene2011}
Stanislas Dehaene and Laurent Cohen.
\newblock {The unique role of the visual word form area in reading}.
\newblock {\em Trends in Cognitive Sciences}, 15(6):254--262, 2011.

\bibitem{McCandliss2003}
Bruce~D. McCandliss, Laurent Cohen, and Stanislas Dehaene.
\newblock {The visual word form area: Expertise for reading in the fusiform gyrus}.
\newblock {\em Trends in Cognitive Sciences}, 7(7):293--299, 2003.

\bibitem{Grill-Spector2014}
Kalanit Grill-Spector and Kevin~S. Weiner.
\newblock {The functional architecture of the ventral temporal cortex and its role in categorization}.
\newblock {\em Nature Reviews Neuroscience}, 15(8):536--548, 2014.

\bibitem{Haxby2001}
J.~V. Haxby, M.~I. Gobbini, M.~L. Furey, A.~Ishai, J.~L. Schouten, and P.~Pietrini.
\newblock {Distributed and overlapping representations of faces and objects in ventral temporal cortex}.
\newblock {\em Science}, 293(5539):2425--2430, 2001.

\bibitem{Cadieu2014}
Charles~F. Cadieu, Ha~Hong, Daniel~L.K. Yamins, Nicolas Pinto, Diego Ardila, Ethan~A. Solomon, Najib~J. Majaj, and James~J. DiCarlo.
\newblock {Deep Neural Networks Rival the Representation of Primate IT Cortex for Core Visual Object Recognition}.
\newblock {\em PLOS Computational Biology}, 10(12):e1003963, 2014.

\bibitem{Cichy2016}
Radoslaw~Martin Cichy, Aditya Khosla, Dimitrios Pantazis, Antonio Torralba, and Aude Oliva.
\newblock {Comparison of deep neural networks to spatio-temporal cortical dynamics of human visual object recognition reveals hierarchical correspondence}.
\newblock {\em Scientific Reports}, 6(1):1--13, 2016.

\bibitem{Kubilius2019}
Jonas Kubilius, Martin Schrimpf, Kohitij Kar, Rishi Rajalingham, Ha~Hong, Najib~J Majaj, Elias~B Issa, Pouya Bashivan, Jonathan Prescott-Roy, Kailyn Schmidt, Aran Nayebi, Daniel Bear, Daniel L~K Yamins, and James~J Dicarlo.
\newblock {Brain-Like Object Recognition with High-Performing Shallow Recurrent ANNs}.
\newblock {\em Advances in Neural Information Processing Systems (NeurIPS)}, 32, 2019.

\bibitem{Yamins2014}
Daniel~L.K. Yamins, Ha~Hong, Charles~F. Cadieu, Ethan~A. Solomon, Darren Seibert, and James~J. DiCarlo.
\newblock {Performance-optimized hierarchical models predict neural responses in higher visual cortex}.
\newblock {\em Proceedings of the National Academy of Sciences of the United States of America}, 111(23):8619--8624, jun 2014.

\bibitem{Lu2023}
Zitong Lu and Yixuan Ku.
\newblock {Bridging the gap between EEG and DCNNs reveals a fatigue mechanism of facial repetition suppression}.
\newblock {\em iScience}, 26:108501, 2023.

\bibitem{Yamins2016}
Daniel~L.K. Yamins and James~J. DiCarlo.
\newblock {Using goal-driven deep learning models to understand sensory cortex}.
\newblock {\em Nature Neuroscience}, 19(3):356--365, 2016.

\bibitem{Margalit2024}
Eshed Margalit, Hyodong Lee, Dawn Finzi, James~J. DiCarlo, Kalanit Grill-Spector, and Daniel~L.K. Yamins.
\newblock {A unifying framework for functional organization in early and higher ventral visual cortex}.
\newblock {\em Neuron}, 112(14):2435--2451.e7, 2024.

\bibitem{Lu2023d}
Zitong Lu and Julie~D Golomb.
\newblock {Human EEG and artificial neural networks reveal disentangled representations of object real-world size in natural images}.
\newblock {\em bioRxiv}, 2023.

\bibitem{Huang2021}
Taicheng Huang, Zonglei Zhen, and Jia Liu.
\newblock {Semantic Relatedness Emerges in Deep Convolutional Neural Networks Designed for Object Recognition}.
\newblock {\em Frontiers in Computational Neuroscience}, 15:625804, 2021.

\bibitem{Guclu2015}
Umut G{\"{u}}{\c{c}}l{\"{u}} and Marcel~A.J. van Gerven.
\newblock {Deep Neural Networks Reveal a Gradient in the Complexity of Neural Representations across the Ventral Stream}.
\newblock {\em Journal of Neuroscience}, 35(27):10005--10014, 2015.

\bibitem{Kietzmann2019}
Tim~C. Kietzmann, Courtney~J. Spoerer, Lynn~K.A. S{\"{o}}rensen, Radoslaw~M. Cichy, Olaf Hauk, and Nikolaus Kriegeskorte.
\newblock {Recurrence is required to capture the representational dynamics of the human visual system}.
\newblock {\em Proceedings of the National Academy of Sciences of the United States of America}, 116(43):21854--21863, 2019.

\bibitem{Lu2023c}
Zitong Lu and Julie~D Golomb.
\newblock {Generate your neural signals from mine: individual-to-individual EEG converters}.
\newblock {\em Proceedings of the Annual Meeting of the Cognitive Science Society 45}, 2023.

\bibitem{Lu2023e}
Zejin Lu, Adrien Doerig, Victoria Bosch, Bas Krahmer, Daniel Kaiser, Radoslaw~M Cichy, and Tim~C Kietzmann.
\newblock {End-to-end topographic networks as models of cortical map formation and human visual behaviour: moving beyond convolutions}.
\newblock {\em arXiv}, 2023.

\bibitem{Baek2021}
Seungdae Baek, Min Song, Jaeson Jang, Gwangsu Kim, and Se~Bum Paik.
\newblock {Face detection in untrained deep neural networks}.
\newblock {\em Nature Communications}, 12(1):1--15, 2021.

\bibitem{Xu2021a}
Shan Xu, Yiyuan Zhang, Zonglei Zhen, and Jia Liu.
\newblock {The Face Module Emerged in a Deep Convolutional Neural Network Selectively Deprived of Face Experience}.
\newblock {\em Frontiers in Computational Neuroscience}, 15:1--12, 2021.

\bibitem{Agrawal2024}
Aakash Agrawal and Stanislas Dehaene.
\newblock {Cracking the neural code for word recognition in convolutional neural networks}.
\newblock {\em PLOS Computational Biology}, 20(9):e1012430, 2024.

\bibitem{Dobs2022}
Katharina Dobs, Julio Martinez, Alexander~J.E. Kell, and Nancy Kanwisher.
\newblock {Brain-like functional specialization emerges spontaneously in deep neural networks}.
\newblock {\em Science Advances}, 8(11):8913, mar 2022.

\bibitem{Blauch2022}
Nicholas~M. Blauch, Marlene Behrmann, and David~C. Plaut.
\newblock {A connectivity-constrained computational account of topographic organization in primate high-level visual cortex}.
\newblock {\em Proceedings of the National Academy of Sciences of the United States of America}, 119:e2112566119, 2022.

\bibitem{Keller2021}
T.~Anderson Keller, Qinghe Gao, and Max Welling.
\newblock {Modeling Category-Selective Cortical Regions with Topographic Variational Autoencoders}.
\newblock {\em arXiv}, 2021.

\bibitem{Prince2024}
Jacob~S. Prince, George~A. Alvarez, and Talia Konkle.
\newblock {Contrastive learning explains the emergence and function of visual category-selective regions}.
\newblock {\em Science Advances}, 10:1776, 2024.

\bibitem{Xu2002}
Fei Xu.
\newblock {The role of language in acquiring object kind concepts in infancy}.
\newblock {\em Cognition}, 85(3):223--250, 2002.

\bibitem{Condry2008}
Kirsten~F. Condry and Elizabeth~S. Spelke.
\newblock {The Development of Language and Abstract Concepts: The Case of Natural Number}.
\newblock {\em Journal of Experimental Psychology: General}, 137(1):22--38, 2008.

\bibitem{Carruthers2002}
Peter Carruthers.
\newblock {The cognitive functions of language}.
\newblock {\em Behavioral and Brain Sciences}, 25(6):657--674, 2002.

\bibitem{Abassi2024}
Etienne Abassi and Liuba Papeo.
\newblock {Category-Selective Representation of Relationships in the Visual Cortex}.
\newblock {\em Journal of Neuroscience}, 44(5), 2024.

\bibitem{Li2024}
Jin Li, Kelly~J. Hiersche, and Zeynep~M. Saygin.
\newblock {Demystifying visual word form area visual and nonvisual response properties with precision fMRI}.
\newblock {\em iScience}, 27(12):111481, 2024.

\bibitem{Poldrack2007}
Russell~A. Poldrack.
\newblock {Region of interest analysis for fMRI}.
\newblock {\em Social Cognitive and Affective Neuroscience}, 2(1):67--70, 2007.

\bibitem{Saxe2006}
Rebecca Saxe, Matthew Brett, and Nancy Kanwisher.
\newblock {Divide and conquer: A defense of functional localizers}.
\newblock {\em NeuroImage}, 30:1088--1096, 2006.

\bibitem{Kanwisher1999}
Nancy Kanwisher, Damian Stanley, and Alison Harris.
\newblock {The fusiform face area is selective for faces not animals}.
\newblock {\em NeuroReport}, 10(1):183--187, 1999.

\bibitem{Duncan2009}
Keith~J. Duncan, Chotiga Pattamadilok, Iris Knierim, and Joseph~T. Devlin.
\newblock {Consistency and variability in functional localisers}.
\newblock {\em NeuroImage}, 46(4):1018--1026, 2009.

\bibitem{Kim2021}
Gwangsu Kim, Jaeson Jang, Seungdae Baek, Min Song, and Se~Bum Paik.
\newblock {Visual number sense in untrained deep neural networks}.
\newblock {\em Science Advances}, 7(1), 2021.

\bibitem{Zhou2022}
Liqin Zhou, Anmin Yang, Ming Meng, and Ke~Zhou.
\newblock {Emerged human-like facial expression representation in a deep convolutional neural network}.
\newblock {\em Science Advances}, 8(12):4383, 2022.

\bibitem{Bao2020}
Pinglei Bao, Liang She, Mason McGill, and Doris~Y. Tsao.
\newblock {A map of object space in primate inferotemporal cortex}.
\newblock {\em Nature}, 583(7814):103--108, 2020.

\bibitem{Coggan2023}
David~D Coggan and Frank Tong.
\newblock {Spikiness and animacy as potential organizing principles of human ventral visual cortex}.
\newblock {\em Cerebral Cortex}, 33(13):8194--8217, 2023.

\bibitem{Huang2022}
Taicheng Huang, Yiying Song, and Jia Liu.
\newblock {Real-world size of objects serves as an axis of object space}.
\newblock {\em Communications Biology}, 5(1):1--12, 2022.

\bibitem{Jain2023}
Nidhi Jain, Aria Wang, Margaret~M. Henderson, Ruogu Lin, Jacob~S. Prince, Michael~J. Tarr, and Leila Wehbe.
\newblock {Selectivity for food in human ventral visual cortex}.
\newblock {\em Communications Biology}, 6(1):1--14, 2023.

\bibitem{Khaligh-Razavi2018}
Seyed-Mahdi Khaligh-Razavi, Radoslaw~Martin Cichy, Dimitrios Pantazis, and Aude Oliva.
\newblock {Tracking the Spatiotemporal Neural Dynamics of Real-world Object Size and Animacy in the Human Brain}.
\newblock {\em Journal of Cognitive Neuroscience}, 30(11):1559--1576, 2018.

\bibitem{Khosla2022}
Meenakshi Khosla, N.~Apurva {Ratan Murty}, and Nancy Kanwisher.
\newblock {A highly selective response to food in human visual cortex revealed by hypothesis-free voxel decomposition}.
\newblock {\em Current Biology}, 32(19):4159--4171.e9, 2022.

\bibitem{Konkle2012a}
Talia Konkle and Aude Oliva.
\newblock {A Real-World Size Organization of Object Responses in Occipitotemporal Cortex}.
\newblock {\em Neuron}, 74(6):1114--1124, 2012.

\bibitem{Konkle2013}
Talia Konkle and Alfonso Caramazza.
\newblock {Tripartite Organization of the Ventral Stream by Animacy and Object Size}.
\newblock {\em Journal of Neuroscience}, 33(25):10235--10242, 2013.

\bibitem{Lu2024a}
Zitong Lu, Yile Wang, and Julie~D. Golomb.
\newblock {Achieving More Human Brain-Like Vision via Human Neural Representational Alignment}.
\newblock {\em arXiv}, 2024.

\bibitem{Lu2024b}
Zitong Lu and Yile Wang.
\newblock {Teaching CORnet Human fMRI Representations for Enhanced Model-Brain Alignment}.
\newblock {\em Cognitive Neurodynamics}, (19):61, 2025.

\bibitem{Geirhos2021}
Robert Geirhos, Kantharaju Narayanappa, Benjamin Mitzkus, Tizian Thieringer, Matthias Bethge, Felix~A. Wichmann, and Wieland Brendel.
\newblock {Partial success in closing the gap between human and machine vision}.
\newblock {\em Advances in Neural Information Processing Systems 34}, 34:23885--23899, 2021.

\bibitem{Bhojanapalli2021}
Srinadh Bhojanapalli, Ayan Chakrabarti, Daniel Glasner, Daliang Li, Thomas Unterthiner, and Andreas Veit.
\newblock {Understanding Robustness of Transformers for Image Classification}, 2021.

\bibitem{Deng2009}
Jia Deng, Wei Dong, Richard Socher, Li~Jia Li, Kai Li, and Li~Fei-Fei.
\newblock {ImageNet: A Large-Scale Hierarchical Image Database}.
\newblock {\em 2009 IEEE Conference on Computer Vision and Pattern Recognition, CVPR 2009}, pages 248--255, 2009.

\bibitem{Radford2021}
Alec Radford, Jong~Wook Kim, Chris Hallacy, Aditya Ramesh, Gabriel Goh, Sandhini Agarwal, Girish Sastry, Amanda Askell, Pamela Mishkin, Jack Clark, Gretchen Krueger, and Ilya Sutskever.
\newblock {Learning Transferable Visual Models From Natural Language Supervision}.
\newblock In {\em Proceedings of the International Conference on Machine Learning (ICML)}, 2021.

\bibitem{Stigliani2015}
Anthony Stigliani, Kevin~S. Weiner, and Kalanit Grill-Spector.
\newblock {Temporal Processing Capacity in High-Level Visual Cortex Is Domain Specific}.
\newblock {\em Journal of Neuroscience}, 35(36):12412--12424, 2015.

\end{thebibliography}

\end{document}